\documentclass[10pt, conference]{IEEEtran}
\IEEEoverridecommandlockouts
\usepackage[utf8]{inputenc}
\usepackage{amsmath}
\usepackage{cite}
\usepackage[utf8]{inputenc}
\usepackage[english]{babel}
\usepackage{setspace}
\usepackage{float}

\usepackage{graphicx}
\def\BibTeX{{\rm B\kern-.05em{\sc i\kern-.025em b}\kern-.08em
    T\kern-.1667em\lower.7ex\hbox{E}\kern-.125emX}}
\begin{document}

\title{DecentRAN: Decentralized Radio Access Network for 5.5G and beyond}

\author{Hao Xu, Xun Liu, Qinghai Zeng, Qiang Li, Shibin Ge, Guohua~Zhou and Raymond Forbes\vspace{-20px}
	\thanks{
        This work has been submitted to the IEEE for possible publication.
        Copyright may be transferred without notice, after which this version may no
        longer be accessible
        
		Hao Xu, Qinghai Zeng, Shibin Ge and Guohua Zhou are with Wireless Network Research Department, Huawei Technologies, Shanghai, China. Emails: \{Hal.Xu; Zengqinghai; Geshibin; Guohua.Zhou\}@huawei.com;
		
		Xun Liu is with Poisson Laboratory, Huawei Technologies, Hangzhou, China. Email: liuxun9@huawei.com 
		
		Qiang Li is with Peng Cheng Lab, Shenzhen, China, Email:liq03@pcl.ac.cn
		
		Raymond Forbes is with Huawei Technologies UK. Email: raymond.forbes@huawei.com 
	}}
\maketitle

\begin{abstract}
Radio Access Network faces challenges from privacy and flexible wide area and local area network access. RAN is limited from providing local service directly due to centralized design of cellular network and concerns of user privacy and data security. 
DecentRAN or Decentralized Radio Access Network offers an alternative perspective to cope with the emerging demands of 5G Non-public Network and the hybrid deployment of 5GS and Wi-Fi in the campus network. 
Starting from Public key as an Identity, independent mutual authentication between UE and RAN are made possible in a privacy-preserving manner. With the introduction of decentralized architecture and network functions using blockchain and smart contracts, DecentRAN has ability to provide users with locally managed, end-to-end encrypted 5G NPN and the potential connectivity to Local Area Network via campus routers. Furthermore, the performance regarding throughput and latency are discussed, offering the deployment guidance for DecentRAN.

\end{abstract}
\begin{IEEEkeywords}
5G, Non Public Network, Radio Access Network, Core Network, Decentralization, Blockchain, Decentralized Infrastructure\vspace{-10px}
\end{IEEEkeywords}

\section{Introduction}
5G has seen a sharp rise since its first 3GPP release in 2018, when Release 15 brings the ground breaking 5G System (5GS) with 5G New Radio (NR) for Radio Access Network (RAN), and 5G Core Network (5GC or CN), which handles all users' sensitive information. The early vision of enhanced Mobile Broadband (eMBB), Ultra-reliable Low Latency Communications (URLLC) and massive Machine-Type Communications (mMTC) were eventually rolled out in later releases of 3GPP R15, R16, and R17. The 5.5G aims to bring more advanced features, such as, Uplink Centric Broadband Communication (UCBC), Real-Time Broadband Communication (RTBC), and Harmonized Communication and Sensing (HCS) in the 5.5G defining R18 releases \cite{Huawei2020}.

Though 5GS has a promising growth in terms of all performance metrics, the architecture of 5GS remains unchanged from its first appearance in the public. 
In fact, the 5GS has suffered from the centralized architecture more than ever, with its limitation preventing 5GS spanning easily over the booming edge connectivity demands \cite{Hilary2022}, leashing the full potential of advanced wireless connectivity \cite{5gacia2019}. On the other hand, the principle of RAN-CN deployment is always inline with privacy regulation, and the RAN was decided to not get hands on users privacy, and no delinquency on user privacy is ever allowed due to fault of RAN equipment. 

RAN, the most charismatic organic body of the Mobile Network (MN), are evolving to an intelligent edge of the communication network with potential to become a regional and local network controller. However, in the current form of RAN, it could not be considered a trusted handler of users data, and it is facing critical challenge on its legal status once the RAN has ability to ``\textit{touch}'' the users' data \cite{5gacia2019}. Hence, it is essential to identify the critical enabler for the feasibility of local data governance regarding RAN in a decentralization scope. 


Meanwhile, RAN has limited support for the mobility across multiple serving nodes and networks \cite{ahmad2020}, as the tracking of a user is not possible for a single RAN node in the roaming scenario, and the mobility management has to be offered at a higher level of the network, e.g., Access and Mobility Function (AMF) entity at the CN. In addition to all existing services that the state-of-the-art RAN offers, there are benefits when the RAN is able to provide decentralized services via the support of decentralized identity and local traffic forwarding of Decentralized Network Functions (DNF), such as, decentralized mobility management, edge endogenous security for User Equipment (UE) and Internet of Things (IoT) and support of decentralized applications (dApps) ecosystem.
\begin{figure}
    \includegraphics[width=0.44\textwidth]{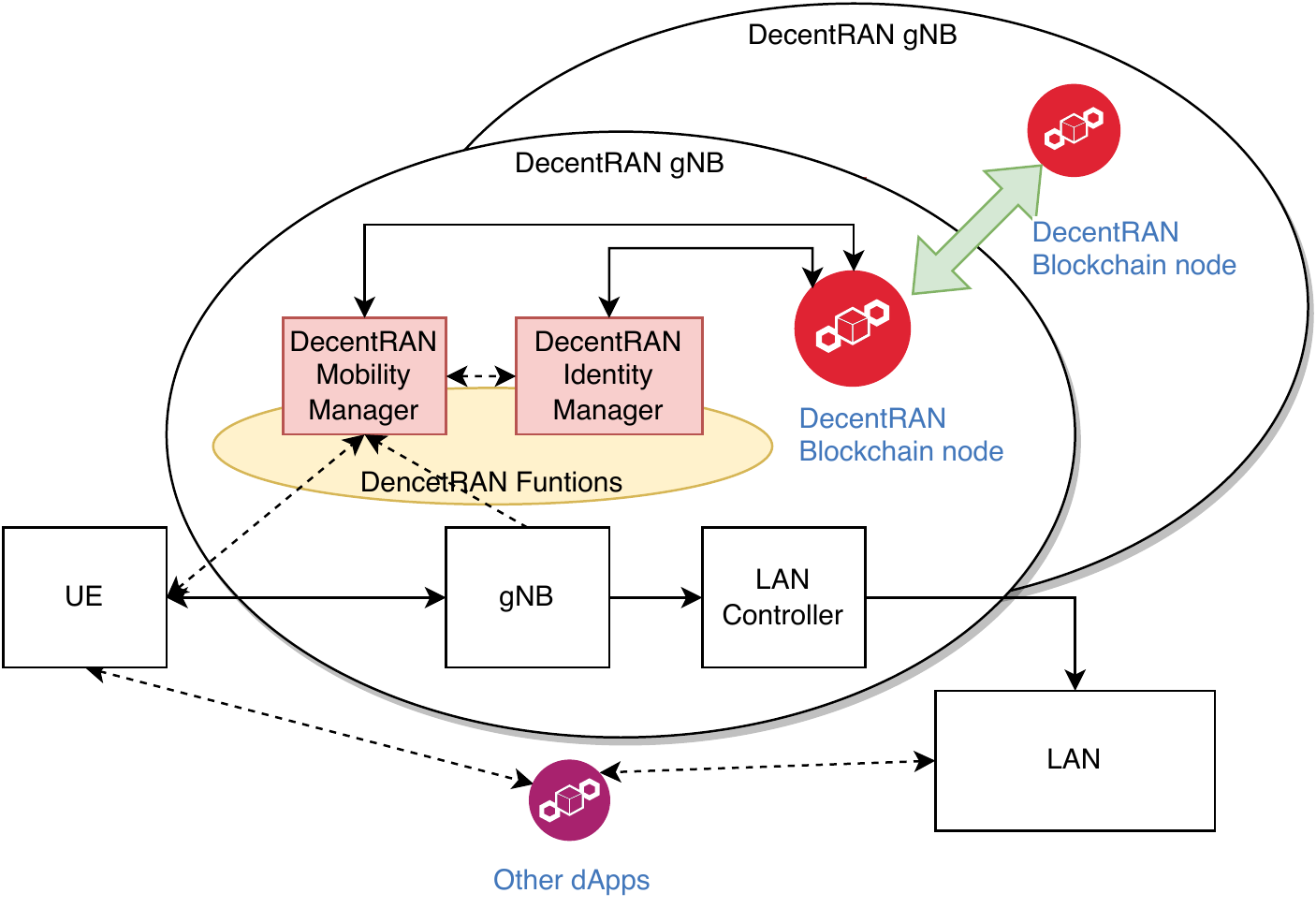}
    \caption{DecentRAN Architecture}
    \vspace{-2em}
    \label{fig:Arch}
\end{figure}
It will be a more secured RAN in the case of Non Public Network, if the identity of user is managed by UE itself so the network can be relieved from threats of privacy leakage and Man-in-the-Middle attacks while providing mobility and network functions. To achieve the synchronization of identity and mobility records, one possible way is employing the distributed ledger technology with asymmetrical cryptography, for all essential identity management, access control, mobility and network functions, as shown in Fig. \ref{fig:Arch}, where the critical user identities are managed under the DecentRAN Identity Manager, which utilizes the decentralized identities from the UE and interact with DecentRAN Mobility Manager for cross gNB handover thanks to the synchronization of mobility context using blockchain.  
On the other hand, the limitation also occurs at the data transfer between peer Next Generation NodeB (gNBs), as they cannot know where user data are going. Therefore, a universal synced identity database is essential for functioning a decentralized RAN infrastructure without revealing any user privacy because of pseudonymity and anonymity of users' identifiers, which is not possible existing Unified Data Management (UDM) deployment without compromising security. 
\vspace{-0.5em}
\subsection{RAN demands for decentralization}
\subsubsection{Lightweight Deployment and Mobility Enhancement}
MNs have spanned through not only the public cellular network, but also popular among industrial and campus networks in which the local user experience is prioritized over wide area network connections. 5G Non public Network (5G NPN) is a novel practice for running local data service over 5G infrastructure with multi-tiers of configurations from sharing the operator managed RAN and CN, to owning the independent RAN and CN with dedicated spectrum \cite{Huawei2021}. Industrial consumers of 5G NPN faces difficult choices choosing from loose privacy and heavy Capital expenditures (CAPEX) and Operational expenditures (OPEX) of owning the standalone 5G infrastructure, as the cost is sensitive to industrial \cite{Hilary2022}, in particular, the Small-Medium-Enterprise (SME). There is a technical gap between current RAN-CN deployment and SME use cases, where the heavy CN raised the bar of entry requirement for 5G NPN solutions.

Local industrial applications require a simplified and light-weighted 5G solution that offers superior mobility for mixed indoor Wireless LAN (WLAN) and outdoor 5G network \cite{5gacia2019}. The network shall also be extensible in the case the communications need to be bridged into the wider area networks, e.g., the Internet.

\subsubsection{Security, Integrity and Privacy Sovereignty driven}
On the other hand, the security and data integrity require the users data to be encrypted from end-to-end, so that all data streams are encrypted while they pass the RAN without inferring the routing service offered by RAN. It brings the dilemma on whether the users' packets can be identified or forwarded based on the network link status and the network identifiers, which could be possibly done by RAN Intelligent Controllers but sacrificing users privacy due to IP/MAC exposure. 
Thereby, if the identity is to be used for data forwarding, it must respect the privacy of users, that's where the decentralized identity with anonymization and pseudonymization kick in. The encryption of identity and data can also be applied to current Local Area Network (LAN) and Wireless Local Area Network (WLAN) practice, together the local network can be run in a fully privacy-preserving, encrypted, and secured manner.  
\vspace{-0.5em}
\subsection{Blockchain and smart contracts integration for decentralized identity and network functions}
Blockchain was widely known as the platform of decentralization since the appearance of Bitcoin in 2008, with its unique combination of consensus and non-repudiate data structure. The blockchain makes the use of decentralized identity, i.e., Public-key as an Identity (PKaaI), a common practice in the decentralized network, and the corresponding decentralized identity management is fully possible thanks to the tamper-proof records stored on the non-repudiate ledger on every single blockchain node. 

The combination of PKaaI and ledger records brings the possibility of decentralized access to RAN in the event users can prove themselves as the rightful holder of the blockchain decentralized identifier (or simply blockchain address, shorten as $\sf BCADD$) with their privileges stated on the blockchain records.  On the other hand, it is straightforward to integrate the decentralized architecture with existing CN and Mobile Network Operator (MNO) management, where they can advertise and insert an BCADD of their own identity to the blockchain, and migrate the trust from the centrally management network. DecentRAN and CN can be played in parallel so that CN can associate central resources to the encrypted identifiers, and Vice Vera. 
Blockchain offers more than just records, smart contracts are supported by major blockchain platform as a Turing-complete automata. It provides the essential functionality for user access control and network activities. 

As blockchain is praised as transparent, tamperproof, privacy-by-design and highly reliable networked system, challenges on performance and scalability, unbalanced privacy and interoperability are still concerning the deployment of blockchain network as a real-time and critical system. 
Further analysis of blockchain security and throughput thresholds can be found in Section \ref{sec:analysis}.


\vspace{-0.5em}
\subsection{Motivations and Contributions}
In the desired local wireless network deployment, there are emerging scenarios require more private and seamless network experience among 5G and other local area wireless networks. It is essential to have the simplified 5GS with loose coupling of CN, while ensuring the same grade of security and privacy protection for the local wireless network users. Therefore we propose Decentralized RAN (DecentRAN) powered by decentralized identity and network functions over blockchain.

\begin{figure}
    \includegraphics[width=0.44\textwidth]{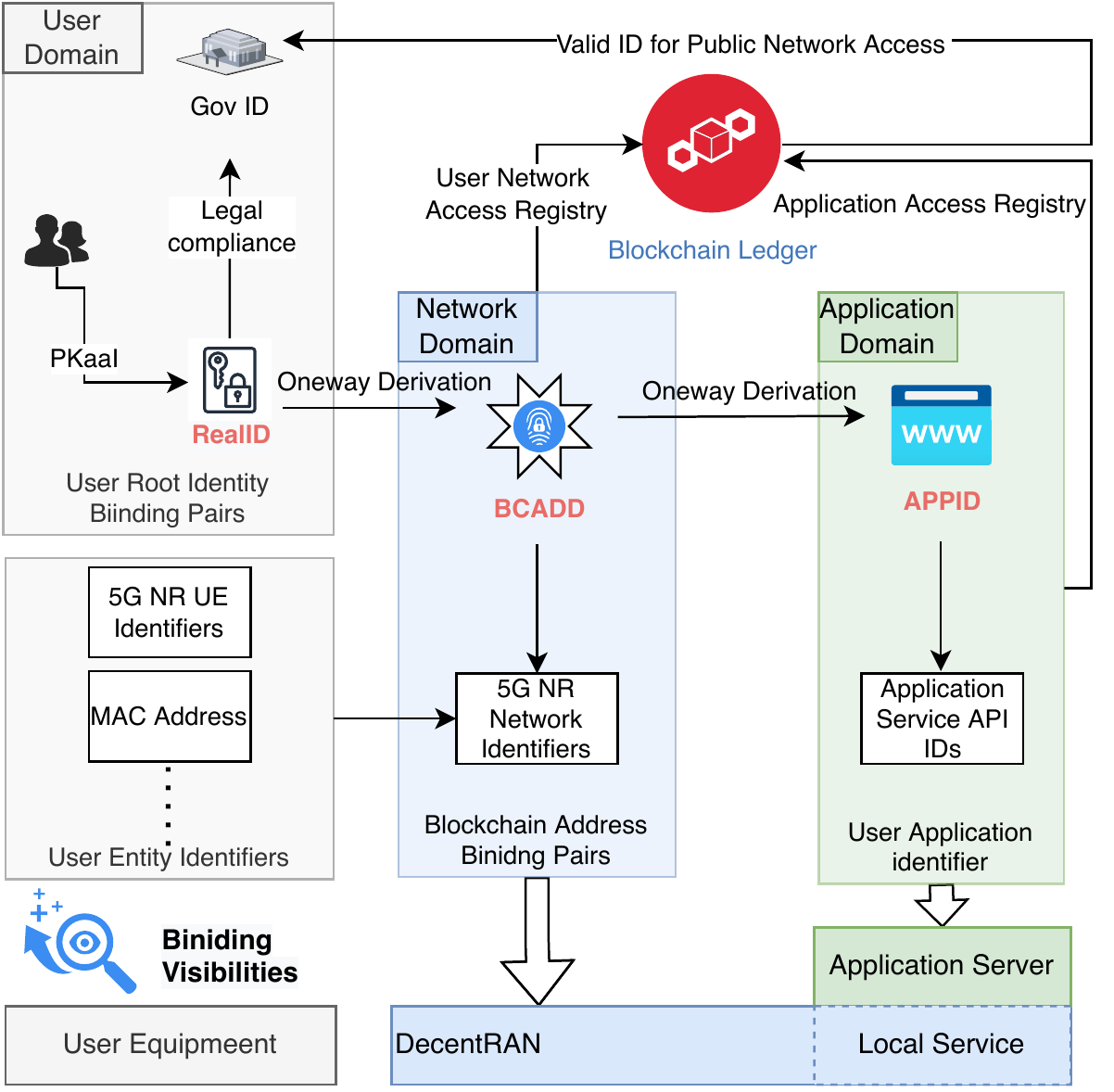}
    \caption{DecentRAN Identity Paradigm }
    \vspace{-2em}
    \label{fig:id}
\end{figure}

This paper contributes to the decentralization of RAN and the corresponding 5GS in four folds. 
\begin{itemize}
    \item We first illustrate an integrated decentralized RAN architecture, described in Fig. \ref{fig:Arch}, for 5GS with decentralized identity, PKaaI, that can be used to look up users and provides necessary cellular network functions, such as initial access with authentication.
    \item Second, we describe decentralization of CN functions and Network dApp based on blockchain and smart contracts, such as Mobility Manager and DNFs, making use of ultra reliable blockchain platform and automating CN requests in a fully decentralized manner.  
    \item Third, we explain how endogenous Security and privacy-preserving are achieved with proposed multi-tier identity architecture, as seen in Fig. \ref{fig:id}, end-to-end encryption (E2EE) and comprehensive synchronizations of public key based identities with status over blockchain, enabling fully encrypted local data forwarding and routing at RAN.
    \item At last, performance analysis of blockchain network for DecentRAN are provided with in house blockchain platform using Raft and HotStuff consensuses, as seen in Fig. \ref{fig:tps} and Fig. \ref{fig:latency}. 
    
\end{itemize}

\section{DecentRAN Architecture}
The decentralized RAN (DecentRAN) is proposed as a novel approach to handle the identity and local traffic in a manner of decentralization for existing User Equipment (UE). It also provides a secured network translation interface, where the WLAN meets 5GS at the LAN Controller, shown in Fig. \ref{fig:Arch}, for inter-domain network traffic. Meanwhile, DecentRAN plays an important role extending the 5GS functionality by fusing application layer services via the blockchain proxy or full blockchain node, offering the user a network managed, self-generated preceding PKaaI identifier for succeeding applications. 
\vspace{-1em}
\subsection{Entities} 
In below, entities of DecentRAN are described in Fig. \ref{fig:Arch}. 

\subsubsection{DecentRAN Identity Manager}

The decentralized identity manager helps decentralized users to initiate mutual authentication with different network entities. The UE initiate Mutual Authentication with serving gNB by using its own {\sf Hash(PK)} (represented as {$\sf BCADD$}) \cite{Xu2021beran}. 


As stated in the PKaaI mechanism, the blockchain address is indicated as {$\sf BCADD$}, which is Tier-2 identity in the three tiers identity infrastructure \cite{Xu2023}.  
Once the UE complete mutual authentication with the network provider, it can further use the Tier-3 identity {$\sf APPID$}, which is derived from {$\sf BCADD$} by using a public function with private parameters for application layer identification.
In the event of regulatory requirements of network identity and legal interception, all entities are required to have the PKaaI identifiers associated with real identity under jurisdiction, and the valid network identifiers shall be verifiable using the {$\sf RealID$} one-way derivable $\sf BCADD$, for legal compliance. Proofs from trusted parties can be carried by users, or accessible from trusted parties APIs. 
\subsubsection{DecentRAN Mobility Manager}
Since all DecentRAN network entities are addressed using $\sf BCADD$, the network is able to look up every one using the decentralized identifier with their topological relations to the RAN and gNBs.  In Mobility Manager, UE is mapped by its global $\sf BCADD$ and the gNB IDs or cell IDs, for the ubiquitous reachability inside and outside network.  The DecentRAN Mobility Manager works in parallel with AMF, UE chooses between CN managed access or the encrypted access.

\subsubsection{Blockchain Node / Proxy and Decentralized Network Functions integration}
Blockchain nodes or proxies and the DNF are mutually hosted, as illustrated in Fig. \ref{fig:Arch}, all DNFs rely on reading and writing to the blockchain ledger, in order to perform mobility and session management, such as mobility manager in Fig. \ref{fig:Arch}. The node or proxy allows the DNF to request the latest blockchain ledger, and writes the updates of UE or RAN states in the blockchain. The proxy can be classified as cache-only proxy, light blockchain node, and full blockchain node. 

For a cache-based read only proxy deployment, DecentRAN has limited functionality on its own, as the network cannot update its latest status to the blockchain network. Therefore, the network is limited performing routing and access control based on local cached information. However, if the mobility updates can be passed on to blockchain nodes via the data interface, DecentRAN will gain mobility support for any committed changes, though DecentRAN cannot manage mobility or access control on its own. 

On the other hand, blockchain can be an integrated part of DecentRAN, where the gNB plays a role of blockchain node. In this case, gNBs forms a consensus group deciding on access policy, mobility using their synchronized ledger in a decentralized manner. The integration of blockchain nodes make the DecentRAN capable of standalone operation with no sensitive information stored or process at any gNBs.
\subsubsection{Network Controller}
The network controller, in the case of NPN, plays the role of LAN controller, as seen in Fig. \ref{fig:Arch}, which is responsible translating $\sf BCADD$ to legacy network addresses, e.g., IP addresses or MAC addresses for legacy network. Meanwhile, the Network Controller is regarded as the combination of UPF and Campus Network Routers if a wider connectivity is needed.


\vspace{-0.5em}
\subsection{Operation model in the era of decentralization }
Given the fact the DecentRAN gNBs are equipped with dedicated decentralized control plane, the operation of DecentRAN may various based on customers requirements and operational guidance of local network market.

\paragraph{Deployment with shared RAN for Multi-operator RAN (MORAN)-- Operator managed gNB}
Enterprise consumer often faces difficult choices if the gNB can only be operated under operator's spectrum. Campus network users have to contract the operator to provide the service to the campus. DecentRAN brings more flexibility on users privacy and data forwarding options, in the case that the gNB has to be run by the operator. DecentRAN allows users to configure their own communication credentials, ensuring the data integrity and security from user domain configurations. 
\paragraph{Standalone Deployment -- Private gNB ownership}
The option of running private gNB usually comes with dedicated spectrum access for the premises spectrum license holder. The user can buy any gNB from any vendors who supports DecentRAN feature. 
\vspace{-0.5em}
\subsection{Case study: Campus network in the case of Logistics and Smart Cities for mixed mobility requirement} 
Private 5G deployment is essential for cutting edge applications of smart cities, IoTs and many emerging industrial use cases, where they demand particular latency, bandwidth, privacy, security and mobility performances. 
To show the advantages of proposed private solution, consistent wireless coverage and seamless network experience are important in modern logistics thanks to massive usages of IoT and personal terminals used by logistics specialists. Meanwhile, in modern day logistic operations, flows of cargo and personnel in mixed indoor and outdoor fields are frequent scenarios in warehouse and in transportation. 

Both indoor and outdoor logistic operations raise challenges for wireless network, as robots, autonomous vehicles and sensors require the network to provide consistent low latency and ultra reliable connections seamlessly while moving around the site. However, the operator faces dilemma on wireless network deployment, as they struggle to choose Wi-Fi or 5G NPN for the mixed indoor and outdoor coverage, and the experience gap brought by mixed Wi-Fi and 5G NPN deployment. The Wi-Fi network is cheap, indoor friendly access technology offer users high speed and low cost connections, on the other hand, 5G base stations offers wider range, more reliable connections, though it is more expensive to own and operate due to the heavy overhead cost of CN. Hence, the logistic wireless network demands for a simplified network deployment that offers benefits of Wi-Fi and 5G with greater coverage and seamless handover between them, in a cost-effective way. 
Similar demands are also found in generic campus networks which involve outdoor and indoor wireless networks. 


\label{sec:analysis}
\vspace{-0.5em}
\subsection{Network security comparisons}
\begin{table*}
    \centering
     \caption{Network Security Comparison}
    \resizebox{0.7\textwidth}{!}{
    \begin{tabular}{c|c|c|c|c}\hline
         Network& Trust model and Trusted parties & Identities & Security Mechanism & Threat Models\\\hline
         5GS & Zoned trust; CN (UE), CA (NE) & SUPI, SUCI, 5G-GUTI
         & 5G-AKA, GBA & MitM\\\hline
         DecentRAN &Minimal Trust; Regulatory & {$\sf RealID$}, {$\sf BCADD$}, {$\sf APPID$} & PKaaI Authentication & MitM, Insider Attack\\\hline
    \end{tabular}
}
\vspace{-1em}
    \label{tab:security}
\end{table*}

Network security has been dramatically switched from centralized symmetrical keys to decentralized asymmetrical keys, the following comparisons reveal how DecentRAN differs from existing 5GS and other networks.

    \subsubsection{Trusted party}  The major difference for DecentRAN and 5GS is who they trust in the first instance. In 5GS, trust is a priori knowledge to all elements with a valid certificate issued by a trusted Certificate Authority (CA) through Operations, Administration and Maintenance (OAM) platform maintained by the MNO. Based on the trusted connections between all RAN and CN elements, 5GS further establishes trust for the users via implementing  UDM/ARPF, which is a fully trusted server with user subscription dataset and credentials. The authentication function server (AUSF) and security anchor function (SEAF) are assembled with other network entities as trusted components to provide security computation to verify users at different level. 
    
    Since the DecentRAN is built on blockchain, it inherits the zero trust (minimal trust) property of blockchain with PKaaI Authentication between any entities among both network elements and UE. An example of UE-gNB PKaaI Mutual Authentication \cite{Xu2021beran}, where the core procedures can be applied in between UE-to-UE, UE-to-Network Equipment (NE), and NE-to-NE. 
    Moreover, by combining the PKaaI mechanism with self-sovereign identity mechanism, the certificate authorization is not necessary in the DecentRAN network for the usage of public key, as the entity choose to trust the blockchain ledger and the associated transactions history of the visiting entity.
    Note that, the blockchain ledger has resistance arbitrary attacks on data, but has no responsibility to the authenticity of data, unless the data can be proven native to the blockchain network, back-traceable to the genesis block. 
    
    \subsubsection{Identity} There are two folds of identities in the network, User domain identities and Network domain identities, as listed in Table \ref{tab:security}. For User Domain identities, 3GPP has defined Subscription Permanent Identifier (SUPI), Subscription Concealed Identifier (SUCI), and compatible 5G Global Unique Temporary Identifier (5G-GUTI)\cite{Zhang2017}. 
    SUCI and 5G-GUTI are derived from SUPI and used for NAS message to resist privacy leakages and achieve mutual authentication. However, the privacy of the SUCI and 5G-GUTI are ensured by the key provided by trusted parties, i.e. CN. Also, the SUPI might be tracked by SUCI-catcher with link-ability of 5G-AKA \cite{Chlosta2021}. 
    
    However, DecentRAN supports three tiers identity framework, which includes {$\sf RealID$},{$\sf BCADD$}, and {$\sf APPID$} for extended application layer identity services. 
    It only requires Regulator as an honest trusted party which holds entities' {$\sf RealID$} and supervises the network when legal compliance of DecentRAN is required, otherwise, the network runs on zero-trust principle.  In the event of legal compliant DecentRAN, the {$\sf RealID$} is never revealed in the network, but {$\sf BCADD$} is required to be derived from the registered {$\sf RealID$}. The {$\sf BCADD$} can not be back linked to {$\sf RealID$} and updated regularly within a secure time slot. By using the PKaaI authentication mechanism\cite{Xu2021beran}, {$\sf BCADD$} and {$\sf APPID$} can achieve mutual authentication and identity mapping as the SUCI and 5G-GUTI do.
    
    
    \subsubsection{Security Architecture} The current 5GS has divided its security domain into three: network domain, user domain, and application domain. The network domain security includes network access security and securely data exchange among network nodes. The fundamental of network domain security is 5G-AKA protocol\cite{Zhang2017}, which enables mutual authentication among network entities and UEs to access or provide service via network securely. The 5G-AKA protocol achieves mutual authentication among the UE, RAN and CN. It is able to ensure confidentiality and integrity over control plane and user plane data exchange. The credential used for 5G-AKA is guarded by user domain security measure with USIM for end-to-end encryption and identification. The application domain security is ensured by the Generic Bootstrap Architecture (GBA) protocol, which is not in the coverage of RAN perspective, but it is related to the later usage of DecentRAN {$\sf APPID$} for the extended application layer services.
    
    For the DecentRAN network, based on the PKaaI Authentication in \cite{Xu2021beran}, a new security architecture which integrates network access authentication and service authorization can be proposed. The network domain security is based on the mutual authentication mechanism, which achieves the same security function compared with 5G-AKA protocol. Based on the design of routing package, DecentRAN can also achieve secure data exchange in network with confidentiality and integrity. For the application domain, its security is also based on the general mutual authentication protocol \cite{Xu2021beran} as well, which achieves mutual authentication on the {$\sf APPID$} with the help of blockchain infrastructure. The {$\sf APPID$} is required to be frequently updated, in order to protect users from {$\sf APPID$}-catcher and tracking of application activities.
    
    The user domain security include the protection of users' public key pairs, their parameters for updating future keys, and the {$\sf RealID$}, {$\sf BCADD$} and {$\sf APPID$}. Users are required to enable password or bio-metric measures to protect the credentials in the UE. Furthermore, the UE is required to enable Trusted Platform Modules (TPM) and Trustable Execution Environment (TEE) to protect keys in a trustworthy manner. 
    \subsubsection{Analysis of adversary models} 
    
    PKaaI Authentication can resist man-in-the-middle attack as the current RAN network does. User's network address ${\sf ADD}$ is bound with {$\sf BCADD$}, and {$\sf BCADD$} is not only an identity identifier but also a one-way hash of the public key. Thus an adversary cannot claim other parties' identity, which is ({$\sf BCADD$}) with his own ${\sf ADD}$ because it breaks the binding relation. The attack is not valid even the adversary receives the message sent to designated {$\sf BCADD$}, as the adversary cannot produce the corresponding private key to decrypt the message. 
    
    In addition, PKaaI provides resistance to insider attacks, as the trust was not a presumption in any circumstance unless they have mutually authenticated each other. Comparing with centralized management and authorization of RAN elements, DecentRAN does not require trusted network or manager to configure its transportation credentials, therefore, it has resistance to known insider attacks. The security of identity is built on the top of blockchain's non-reputable and secured ledger, where the security is ensured by distributed consensus against arbitrary attacks within security threshold.

\vspace{-0.3em}
\section{Results and Discussions}
DecentRAN performance is simulated using HuaweiChain hosted by Huawei Cloud Blockchain Services \cite{Huawei2022} with 4 peer nodes and 4 consensus nodes running on Intel Xeon Gold 6266C at 3.00 GHz, each node is allocated with 16 cores, 64 GB RAM and Network Accessible Storage Reading and Writing at 350MB/s respectively. Each node represents a blockchain node which are shared among certain number of gNBs by load balancing and fault tolerance policies. HuaweiChain offers a hybrid peer configurations, potentially consensus nodes can be selected dynamically from all participating RAN nodes for improved scalability and reliability. In HuaweiChain, peer nodes are categorized into validator and consensus nodes, where all transactions are sorted and packed into blocks by the featured consensuses, which are Raft, Hotstuff and Solo (only engages one node for comparison purpose). Thanks to the plug-n-play consensus module, multiple consensus mechanisms can also be added into the network for versatile choices between Crash Fault Tolerance and Byzantine Fault Tolerance scenarios.

\subsection{Inter-gNB throughput study}
\begin{figure}[h]
    \includegraphics[width=0.44\textwidth]{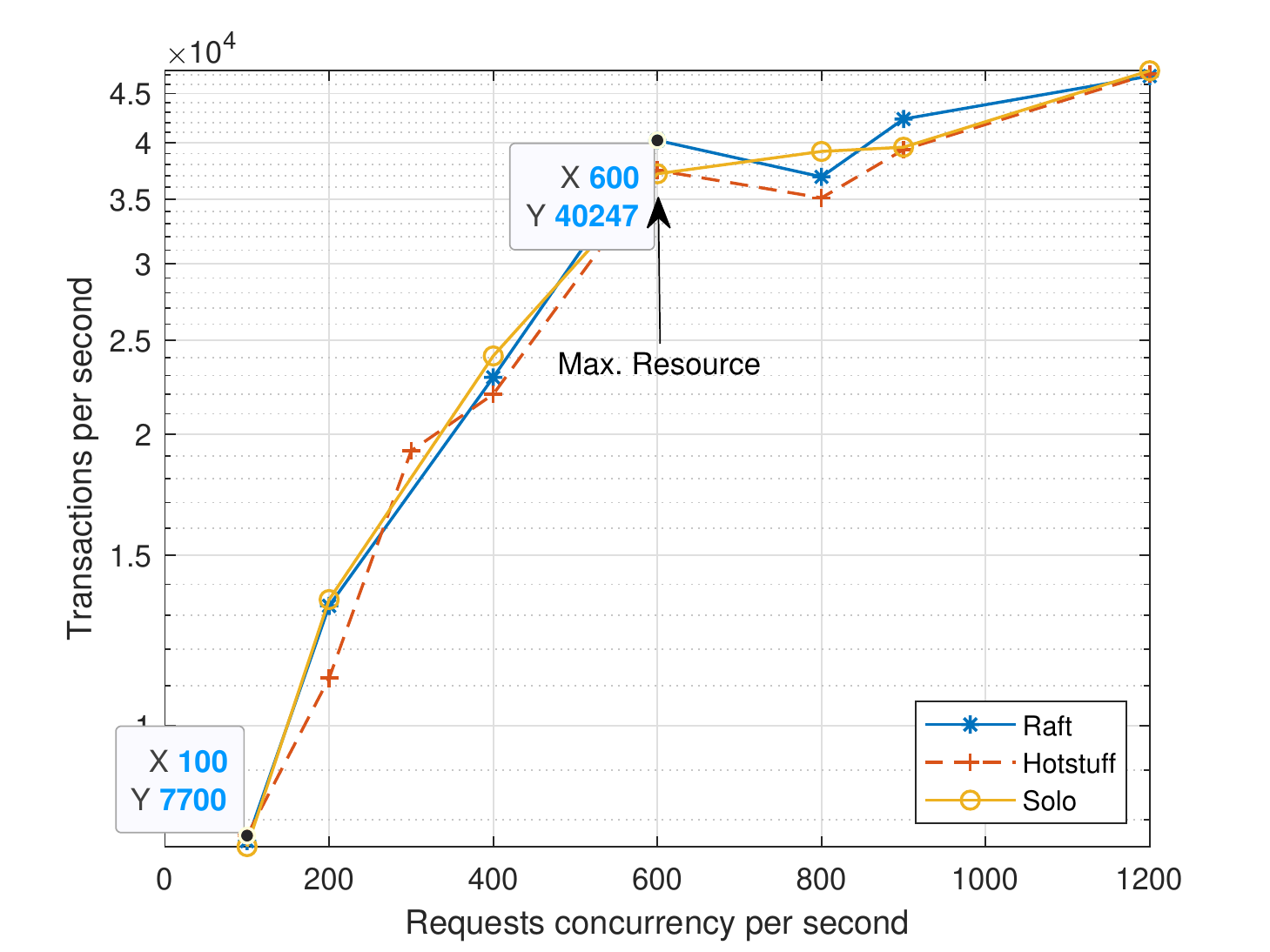}
    \caption{Transactions per second vs concurrent requests}\vspace{-0.1em}
    \label{fig:tps}
\end{figure}

In Fig. \ref{fig:tps}, throughput of Raft and Hotstuff are compared with Solo by feeding concurrent requests in steps. All requests are sent by 6 identical servers as the serving node, with 200 Byte per transaction. Client network has 10,000 Mbps bandwidth with average latency of 0.2ms. All networks throughput increases when requests are rising, and reaches the maximum top around 40000 Transactions per second, when the CPU capacity is exhausted. 

\vspace{-0.5em}
\subsection{Inter-gNB latency study}
\begin{figure}
    \includegraphics[width=0.44\textwidth]{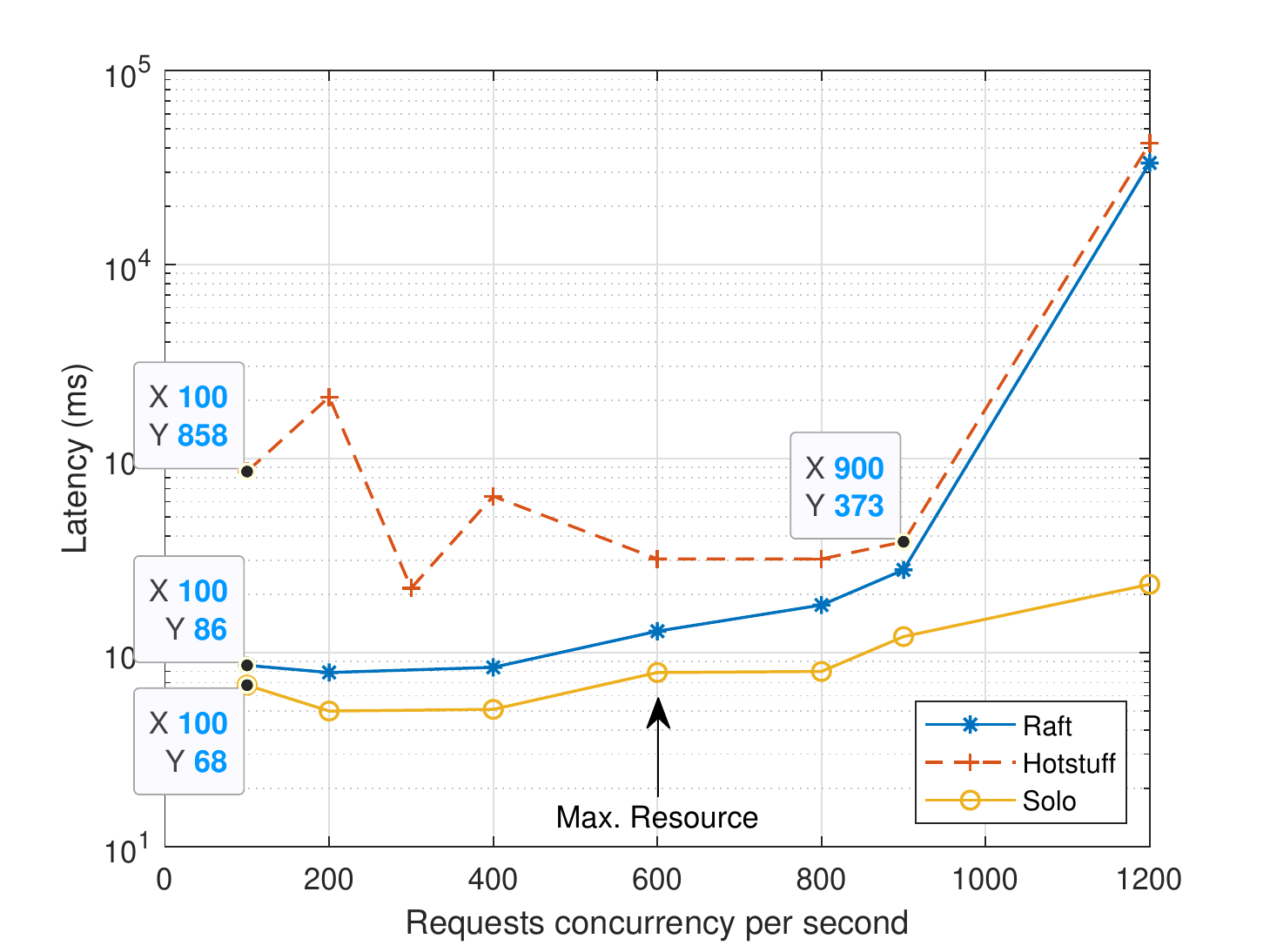}
    \caption{Latency vs concurrent requests}
    \label{fig:latency}
    \vspace{-2em}
\end{figure}

In the latency simulation, shown in Fig. \ref{fig:latency}, Raft and Solo have rather progressive result when the network is within the resource capacity, maintaining the E2E latency around 100ms, and increases dramatically when the network is overloaded. On the other hand, Hotstuff has significant delay when the network is not fully fed by the traffic, as the hotstuff adopts a leader rotation mechanism, which only packs the block when the leader rotation is occurred, similar to the TPS performance drop at earlier simulations on throughput.  

The results provide a glimpse on how well the ordinary consensus work on a universal middle-range computing hardware. The results suggest the initial performance on writing and conducting decentralized services on DecentRAN are satisfactory for non real time initial registration, mobility update and other advanced network functions that require writing on blockchain. Meanwhile, it is worth-noting that most operations, such as authentication, handover based on the current mobility status, data forwarding and routing are in real time, as the data required are cached on the local blockchain node or proxy.
\vspace{-0.5em}
\subsection{Challenges}
The overall performance offers the E2E RAN operations with 40,000 writes per second in around 100ms delay. It is hard to declare the total victory of decentralized RAN operation against the existing centralized solution with mediocre latency performance, though the system proves itself to be highly scalable with impressive overall throughput. On the other hand, it is certain that the latency can be improved by scarifying throughput, as the real time block generation and distribution with less transactions in one block would consume more resources in general. 
Consequently, it is critical to evolve the blockchain platform \cite{Peng2022} with better ordering and validating designs \cite{Qi2021} to boost the latency of DecentRAN without limiting the growth of throughput, as followed by the future work of DecentRAN.


\section{Conclusions}\vspace{-0.3em}
In the desired local wireless network deployment, there are emerging scenarios require more private and seamless network experience among 5G and other local area wireless networks. Therefore, it is essential to have the simplified 5GS with loose coupling of CN, while ensuring the same grade of security and privacy protection for the local wireless network users. In this paper, we proposed DecentRAN powered by decentralized identity and network functions over blockchain and smart contracts, simplifying core deployment and securing 5GS at edge and campus. We also studied the security aspects of DecentRAN with universal authentications for networks and applications. Moreover, the performances of DecentRAN throughput and latency are investigated, showing promising results of existing consensus, and indicating the potential of improvements for bespoke real time consensus.  

In future work, it is necessary to seek novel and lightweight blockchain consensuses that offer adequate latency and throughput with dedicated consideration of wireless communication networks. The future work aims to provide RAN with native consensus communication stacks on RAN Radio Resource Control messages and interoperability between Mobile Network and Data Network.

\bibliographystyle{IEEEtran}
\bibliography{references}

\begin{thebibliography}{10}
\providecommand{\url}[1]{#1}
\csname url@samestyle\endcsname
\providecommand{\newblock}{\relax}
\providecommand{\bibinfo}[2]{#2}
\providecommand{\BIBentrySTDinterwordspacing}{\spaceskip=0pt\relax}
\providecommand{\BIBentryALTinterwordstretchfactor}{4}
\providecommand{\BIBentryALTinterwordspacing}{\spaceskip=\fontdimen2\font plus
\BIBentryALTinterwordstretchfactor\fontdimen3\font minus
  \fontdimen4\font\relax}
\providecommand{\BIBforeignlanguage}[2]{{%
\expandafter\ifx\csname l@#1\endcsname\relax
\typeout{** WARNING: IEEEtran.bst: No hyphenation pattern has been}%
\typeout{** loaded for the language `#1'. Using the pattern for}%
\typeout{** the default language instead.}%
\else
\language=\csname l@#1\endcsname
\fi
#2}}
\providecommand{\BIBdecl}{\relax}
\BIBdecl

\bibitem{Huawei2020}
\BIBentryALTinterwordspacing
{Huawei Technologies}, ``{Huawei's David Wang: Defining 5.5G for a Better,
  Intelligent World},'' Online, 2020. [Online]. Available:
  \url{https://www.huawei.com/en/news/2020/11/mbbf-shanghai-huawei-david-wang-5dot5g}
\BIBentrySTDinterwordspacing

\bibitem{Hilary2022}
H.~Frank, C.~Colman-Meixner, K.~D.~R. Assis, S.~Yan, and D.~Simeonidou,
  ``{Techno-Economic Analysis of 5G Non-Public Network Architectures},''
  \emph{IEEE Access}, vol.~10, pp. 70\,204--70\,218, 2022.

\bibitem{5gacia2019}
{5G Alliance for Connected Industries and Automation}, ``{5G Non-Public
  Networks for Industrial Scenarios},'' 2019.

\bibitem{ahmad2020}
R.~Ahmad, E.~A. Sundararajan, and A.~Khalifeh, ``{A survey on femtocell
  handover management in dense heterogeneous 5G networks},''
  \emph{Telecommunication Systems}, vol.~75, pp. 481--507, 2020.

\bibitem{Huawei2021}
\BIBentryALTinterwordspacing
Huawei, ``{Xinyan Coal Mine, Shanxi Mobile, and Huawei Launch 5GtoB PNI-NPN
  Kite-like SolutionXinyan Coal Mine, Shanxi Mobile, and Huawei Launch 5GtoB
  PNI-NPN Kite-like Solution},'' Online, 2021. [Online]. Available:
  \url{https://www.huawei.com/en/news/2021/2/kite-like-solution}
\BIBentrySTDinterwordspacing

\bibitem{Xu2021beran}
\BIBentryALTinterwordspacing
H.~Xu, L.~Zhang, Y.~Sun, and C.-L. I, ``{BE-RAN: Blockchain-enabled Open RAN
  with Decentralized Identity Management and Privacy-Preserving
  Communication},'' jan 2021. [Online]. Available:
  \url{http://arxiv.org/abs/2101.10856}
\BIBentrySTDinterwordspacing

\bibitem{Xu2023}
H.~Xu, Y.~Sun, Z.~Li, Y.~Sun, X.~Zhang, and L.~Zhang, ``{deController: A Web3
  Native Cyberspace Infrastructure Perspective},'' \emph{IEEE communications
  magazine}, 2023.

\bibitem{Zhang2017}
X.~Zhang, A.~Kunz, and S.~Schroder, ``{Overview of 5G security in 3GPP},'' in
  \emph{2017 IEEE Conference on Standards for Communications and Networking
  (CSCN)}.\hskip 1em plus 0.5em minus 0.4em\relax IEEE, sep 2017, pp. 181--186.

\bibitem{Chlosta2021}
M.~Chlosta, D.~Rupprecht, C.~P{\"{o}}pper, and T.~Holz, ``{5G SUCI-Catchers:
  Still Catching Them All?}'' in \emph{Proceedings of the 14th ACM Conference
  on Security and Privacy in Wireless and Mobile Networks}, ser. WiSec
  '21.\hskip 1em plus 0.5em minus 0.4em\relax New York, NY, USA: Association
  for Computing Machinery, 2021, pp. 359--364.

\bibitem{Huawei2022}
\BIBentryALTinterwordspacing
{Huawei Technologies}, ``{Huawei Cloud Blockchain Services},'' 2022. [Online].
  Available: \url{https://www.huaweicloud.com/intl/en-us/product/bcs.html}
\BIBentrySTDinterwordspacing

\bibitem{Peng2022}
Z.~Peng, Y.~Zhang, Q.~Xu, H.~Liu, Y.~Gao, X.~Li, and G.~Yu, ``{NeuChain: A Fast
  Permissioned Blockchain System with Deterministic Ordering},'' \emph{Proc.
  VLDB Endow.}, vol.~15, no.~11, pp. 2585--2598, 2022.

\bibitem{Qi2021}
J.~Qi, X.~Chen, Y.~Jiang, J.~Jiang, T.~Shen, S.~Zhao, S.~Wang, G.~Zhang,
  L.~Chen, M.~H. Au, and H.~Cui, \emph{{Bidl: A High-Throughput, Low-Latency
  Permissioned Blockchain Framework for Datacenter Networks}}.\hskip 1em plus
  0.5em minus 0.4em\relax New York, NY, USA: Association for Computing
  Machinery, 2021, pp. 18--34.

\end{thebibliography}

\end{document}